\documentclass{PoS}

\newcommand{\beq}{\begin{equation}}
\newcommand{\eeq}{\end{equation}}
\newcommand{\ba}{\begin{eqnarray}}
\newcommand{\ea}{\end{eqnarray}}

\newcommand{\bm}{\mathbf}

\newcommand{\simorder}{\raisebox{-4pt}{$\, \stackrel{\textstyle >}{\sim} \,$}}

\title{Linearly Polarized Gluon Effects in Unpolarized Collisions}

\ShortTitle{Linearly Polarized Gluon Effects in Unpolarized Collisions}

\author{\speaker{Dani\"el Boer}\thanks{I thank John Collins, Markus Diehl, and Ted Rogers for helpful discussions, Wilco den Dunnen, Piet Mulders, Cristian Pisano, Marc Schlegel, Werner Vogelsang for fruitful collaborations, and Miguel Echevarr\'ia for providing me with results of reference \cite{Echevarria:2015uaa} and feedback.}\\
        Van Swinderen Institute for Particle Physics and Gravity, University of Groningen\\ 
Nijenborgh 4, NL-9747 AG Groningen, The Netherlands\\
        E-mail: \email{d.boer@rug.nl}}

\abstract{Linear polarization of gluons inside unpolarized hadrons affects the transverse momentum distribution of produced spin-0 particles, such as of the Higgs or (pseudo-)scalar quarkonium states at LHC. Despite the currently unknown amount of linear gluon polarization, a range of predictions can be obtained, using TMD evolution, which indicates that their effect is on the few percent level in Higgs production, but can be much larger in quarkonium production. Together with asymmetries in open charm or bottom production in electro-production at an Electron-Ion Collider, the size and sign of the linear gluon polarization could be extracted experimentally. These processes also allow to test the behavior expected at small $x$ in and outside the saturation region.}

\FullConference{QCD Evolution 2015\\
                 May 26-30, 2015\\
                 Jefferson Lab (JLAB), Newport News Virginia, USA}

\begin{document}

\section{Introduction}
Gluons inside unpolarized hadrons can be linearly polarized \cite{Mulders:2000sh}. It corresponds to an interference between 
$\pm 1$ gluon-helicity states and requires nonzero transverse momentum. The distribution of linearly polarized gluons inside 
an unpolarized hadron is thus described by a transverse momentum dependent distribution function (TMD), here denoted by 
$h_1^{\perp\, g}$. For $h_1^{\perp\, g} > 0$ the gluon polarization $\epsilon_T$ is preferentially along $k_T$, with a $\cos 2\phi$ distribution 
around it, where $\phi = \angle (k_T,\epsilon_T)$. This TMD is $k_T$-even, chiral-even and T-even, and is fully allowed 
by the symmetries of QCD. Linear polarization of gluons is generated perturbatively, i.e. by radiative corrections, which was 
first noted in a study of $p\,p\to \gamma \,\gamma \, X$ \cite{Nadolsky:2007ba}. It modifies the transverse momentum distribution of Higgs production 
($\sigma(Q_T)$) at NNLO pQCD \cite{Catani:2010pd,Wang:2012xs}. Nonperturbatively it can be present at tree level, affecting 
Higgs production at low $Q_T$ \cite{Sun:2011iw,Boer:2011kf}. In this overview of the effects of linear gluon polarization in unpolarized collisions, 
first the transverse momentum distribution of Higgs production is discussed. Estimates are given using TMD evolution. 
Subsequently, (pseudo-)scalar $C=+$ quarkonium production is studied, heavy quark pair and dijet production are discussed, and the expectations in the small-$x$ limit are commented on. 
	 
\section{Higgs transverse momentum distribution}
\begin{figure}[htb]
\centerline{
\includegraphics[width=6.5 cm]{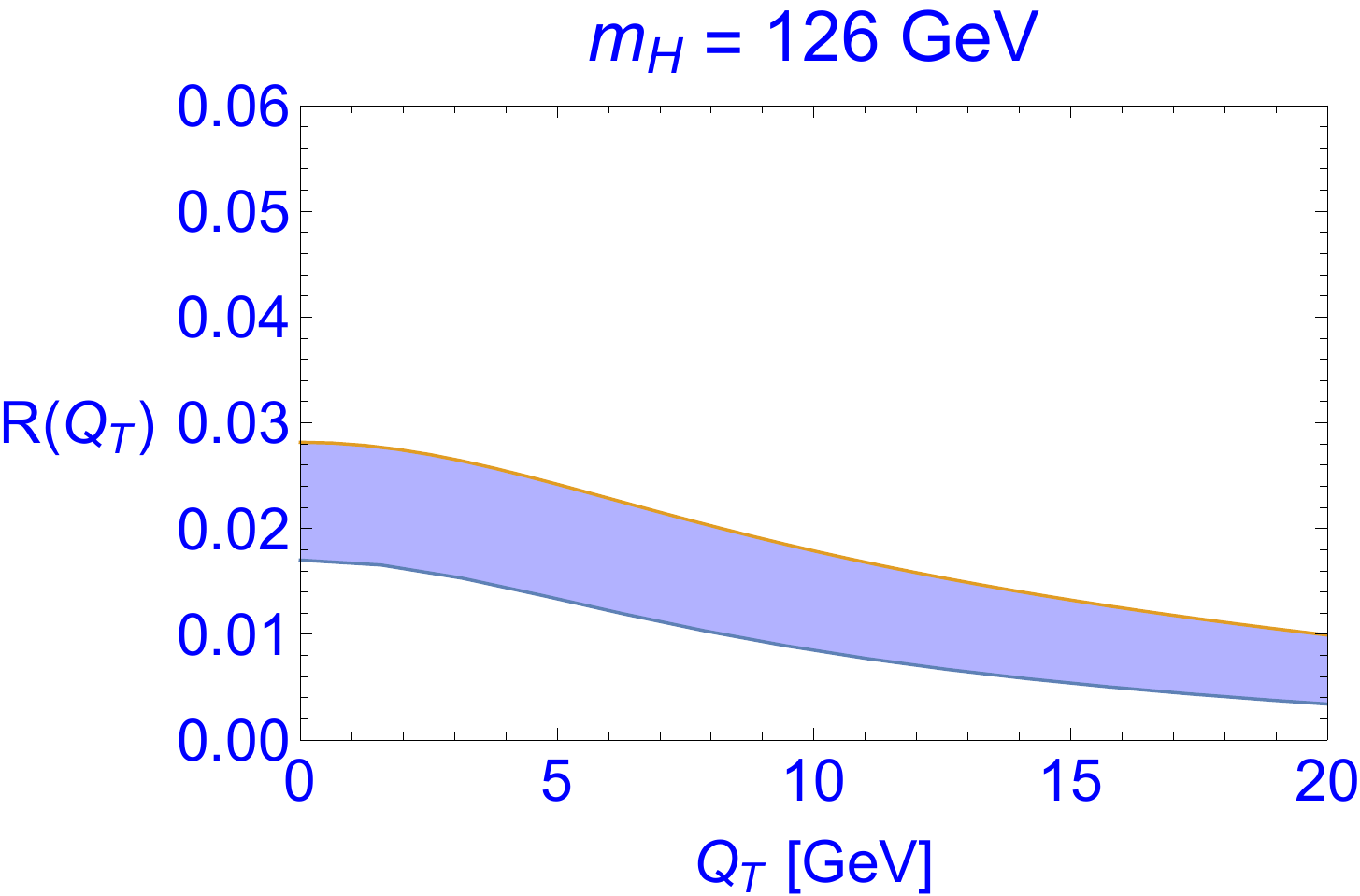}\hspace{1 mm}
\includegraphics[width=6.5 cm]{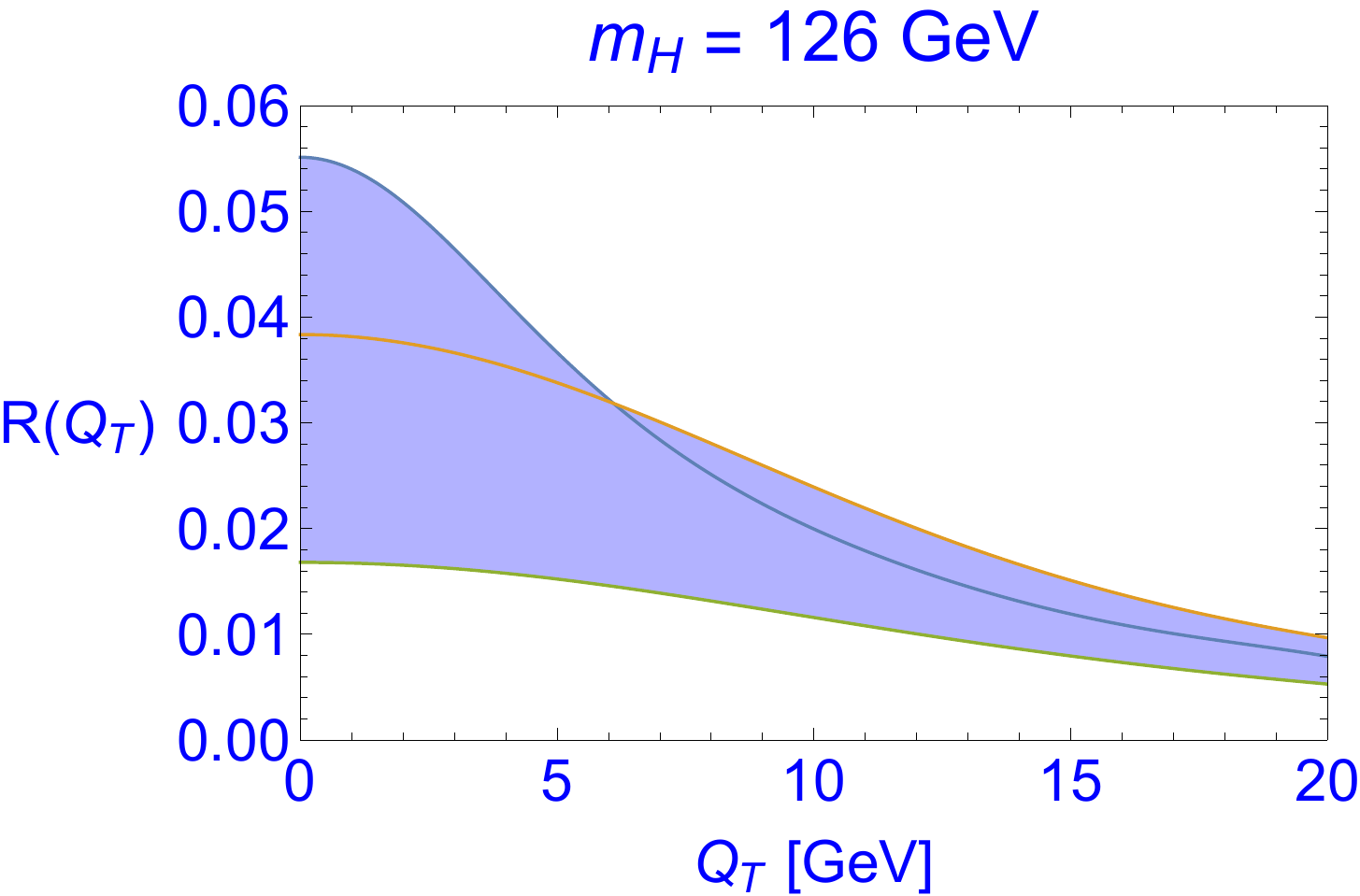}
}
\caption{\label{RQTatHiggs}Range of predictions for ${\cal R}(Q_T)$ of \cite{Boer:2014tka} (left) and of \cite{Echevarria:2015uaa} (right).} 
\end{figure}

Higgs production happens predominantly via gluon fusion $gg\to H$. The inclusive Higgs production cross section is described using collinear factorization (at ${\cal O} (\alpha_s^2)$, i.e.\ NNLO, in \cite{Harlander:2002wh,Anastasiou:2002yz}) and involves only unpolarized collinear parton distributions.
Also the transverse momentum distribution at large $Q_T$ is described by collinear factorization at fixed order in perturbation theory (at NNLO in \cite{deFlorian:1999zd,Ravindran:2002dc,Glosser:2002gm}), but at smaller $Q_T$ large logs of $Q_T/m_H$ need to be resummed. The perturbative state-of-the-art description is with NNLL resummation \cite{deFlorian:2000pr,deFlorian:2001zd,Bozzi:2005wk,Catani:2010pd,deFlorian:2011xf,Becher:2012yn,Neill:2015roa}). At small $Q_T$, which in Higgs production can still be of the order of several GeV, nonperturbative contributions need to be included. In this region it is more natural to describe the process in terms of transverse momentum dependent distributions (TMDs). The TMD factorization expression for $p p \to H X$ has the form \cite{Collins:2011zzd,Echevarria:2015uaa}: 
\beq
d \sigma = {H \times {\rm convolution}\ {\rm of}\ A\, B} + {\rm high}\ q_T\ {\rm correction}\ ({\rm Y}\ {\rm term})+ {\rm power}\ {\rm suppressed}.
\eeq
Here $H$ is the hard partonic scattering factor, and $A$ and $B$ are TMDs, which apart from the transverse momentum also depend on $x$, a rapidity variable $\zeta$ and the renormalization scale $\mu$. The convolution in terms of $A$ and $B$ can be 
deconvoluted by Fourier transforming, which leads to an expression:
\beq
\frac{d\sigma}{dx_A dx_B d\Omega d^2 \bm{q}_{T}^{}} = \int d^2 {b} \, e^{-i {\bm{b} \cdot \bm{q}_T^{}}} \widetilde{W}({\bm{b}}, Q;
  x_a, x_B) + {\cal O}\left(Q_T/Q\right), \label{diffxs}
\eeq
where the hard scale $Q$ is set by the Higgs mass $m_H$ and $Q_T^2 = \bm{q}_T^2$. 
For unpolarized hadrons and unpolarized gluons, $\widetilde W$ for the $gg\to H$ subprocess is given by ($b= |\mathbf{b}|)$
\begin{equation}
\widetilde{W}({b}, Q; x_A, x_B) = H\left(Q\right) \, e^{-S_A(b,Q)} \widetilde{f}_1^g(x_{A},b^2; \mu_b^2, \mu_b) \, 
\widetilde{f}_1^g(x_{B},b^2;\mu_b^2, \mu_b). 
\end{equation}
Here $H\left(Q \right) \propto \left(1 + \alpha_s(Q^2) F_1 + {\cal O}(\alpha_s^2) \right)$, with $F_1$ a renormalization-scheme-dependent finite term. $H(Q)$ contains no large logarithms. The Fourier transformed TMDs $\widetilde{f}_1^g$ are evolved to a $b$-dependent scale  $\mu_b=b_0/b=2e^{-\gamma_E}/b$ $(b_0 \approx 1.123)$.
This means that even at one given value of $Q$ one probes TMDs over a whole range of scales. The Sudakov factor $S_A$ resums large logs in $bQ$. 
The integral over all $b$ includes large, i.e.\ nonperturbative, $b$ values. For this reason one defines $\widetilde{W}(b) \equiv \widetilde{W}(b_*) \, e^{-{S_{NP}(b)}}$, 
with $b_*=b/\sqrt{1+b^2/b_{\max}^2} \leq b_{\max}$. For the common choice $b_{\max} = 1.5\ {\rm GeV}^{-1}$, $\alpha_s(b_0/b_{\max}) = 0.62$, such that $W(b_*)$ can be calculated perturbatively for all $b$ values, but the nonperturbative Sudakov factor $S_{NP}$ cannot. In general $S_{NP}$ is $Q$ dependent. No extraction of $S_{NP}$ for gluons exists yet. As a first guess, one can modify an $S_{NP}$ for quarks from Drell-Yan fits, e.g.\ from \cite{AR}, by rescaling with the color factor ratio $C_A/C_F$ \cite{Boer:2014tka}. 
Finally, polarization of gluons can be included by using the following anisotropic gluonic correlator
\beq
\widetilde \Gamma_g^{ij}(x,\mathbf{b} ) = \frac{1}{2x}\,\bigg \{\delta^{ij}\,\widetilde{f}_1^g(x,b^2)-\bigg( \frac{2b^i b^j}{b^2}\, {-}\, \delta^{ij}\bigg) \;\widetilde{h}_1^{\perp\,g}(x,b^2) \bigg \}  , 
\eeq
which adds a term to $\widetilde W(\mathbf{b})$. In principle, the $Q$-independent part of $S_{NP}$ can be different for the additional term, but that will only have a minor effect at high $Q$ and will be neglected here. Assembling all this, the cross section takes the form \cite{Boer:2011kf}
\beq
\frac{E\,d\sigma^{p p \to H X}}{d^{3}\vec{q}}\Big|_{q_{T}\ll m_{H}} \propto \left(\mathcal{C}\left[f_{1}^{g}\, f_{1}^{g}\right]+\mathcal{C}\left[w_{H}\, h_{1}^{\perp g}\, h_{1}^{\perp g}\right]\right)\,+\mathcal{O}\left(\frac{q_{T}}{m_{H}}\right),
\eeq
where $w_{H}=\left((\mathbf k_{1T}\cdot\mathbf k_{2T})^{2}-\frac{1}{2}\mathbf k_{1T}^{2} \mathbf k_{2T}^{2}\right)/2M^4$ and the (angular independent) relative effect of linearly polarized gluons is given by:
\begin{eqnarray}
{\cal R}(Q_T) & \equiv & \frac{\mathcal{C}[w_H\,h_1^{\perp g}\,h_1^{\perp g}]}{\mathcal{C}[f^g_1\,f_1^g]} \nonumber \\
& = & \frac{\int d^2 \mathbf{b} \, e^{i \mathbf{b} \cdot \mathbf{q}_T^{}} e^{-S_A(b_*,Q) - S_{NP}(b,Q)}
\; \widetilde{h}_1^{\perp g}(x_{A},b_*^2; \mu_{b_*}^2,\mu_{b_*}) \; \widetilde{h}_1^{\perp g}(x_{B},b_*^2;  \mu_{b_*}^2,\mu_{b_*})}{
\int d^2 \mathbf{b} \, e^{i \mathbf{b} \cdot \mathbf{q}_T^{}} \, e^{-S_A(b_*,Q)- S_{NP}(b,Q)} \widetilde{f}_1^g(x_{A},b_*^2; \mu_{b_*}^2,\mu_{b_*}) \, 
\widetilde{f}_1^g(x_{B},b_*^2; \mu_{b_*}^2,\mu_{b_*})}, \label{calRQT}
\end{eqnarray}
where 
\[
\widetilde{h}_1^{\perp\,g}(x,b^2) = \int d^2\mathbf k_T^{}\; \frac{(\mathbf{b}\!\cdot \!
\mathbf k_T^{})^2 - \frac{1}{2}\mathbf{b}^{2} \mathbf k_T^{2}}{b^2 M^2}
\; e^{-i \mathbf{b} \cdot \mathbf{k}_T^{}}\; h_1^{\perp g}(x,k_T^2) =  -\pi \int dk_T^2
\frac{k_T^2}{2M^2} J_2(bk_T) h_1^{\perp g}(x,k_T^2).
\]
Due to the appearance of $b_*$ in the ${\cal R}(Q_T)$ expressions, only the TMDs at small $b$ values are required, the so-called perturbative tails of the TMDs. In leading perturbative order these are given by \cite{Nadolsky:2007ba,Catani:2010pd,Sun:2011iw}
\begin{eqnarray}
\widetilde{f}_1^g(x,b^2; \mu_b^2,\mu_b) & = & f_{g/P}(x; \mu_b) +  {\cal O} (\alpha_s), \nonumber\\
\widetilde{h}_1^{\perp g}(x,b^2; \mu_b^2,\mu_b) & =  & \frac{\alpha_s(\mu_b) C_A}{2\pi}
\int_x^1 \frac{d\hat x}{\hat x} \left(\frac{\hat x}{x}-1\right) f_{g/P}(\hat x; \mu_b) \label{tails}
+ {\cal O} (\alpha_s^2), \label{tailh1perp}
\end{eqnarray}
where $f_{g/P}(x;\mu)$ is the collinear gluon distribution. 
It should be mentioned that although small $b$ roughly corresponds to large $k_T$, the large-$k_T$ behavior can be misleading. Under evolution, TMDs develop a power-law tail at large $k_T$, which can be calculated perturbatively. For both $f_1^g$ and $h_1^{\perp \, g}$ it is $\propto \alpha_s P_{f/h}\otimes f_{g/P}/k_T^2$ for some splitting functions $P_f$ and $P_h$, respectively. The amount of linear gluon polarization at large $k_T$, then becomes $(\alpha_s P_h \otimes f_{g/P})/(\alpha_s P_f \otimes f_{g/P})$ which is not small (50\% or more for $k_T \simorder 10 $ GeV) \cite{Boer:2013fca}. Hence, the ratio of the large-$k_T$ tails of $h_1^{\perp \, g}$ and $f_1^g$ is large, but that does not mean large polarization effects arise in Higgs production at large $Q_T$, because that involves ratios of {\it integrals} over all $k_T$. What matters is the small-$b$ behavior of the Fourier transformed TMDs, Eq.\ (\ref{tails}). The linear polarization at small $b$ starts at order $\alpha_s$, which gives it an unavoidable suppression w.r.t.\ $f_1^g$. 
Numerical analysis in \cite{Boer:2014tka} shows (Fig.\ \ref{RQTatHiggs} (left)) that in Higgs production, linear gluon polarization contributes at the few \% level (at the same level as a fixed order expression at order $\alpha_s(m_H)^2 = {\cal O}(10^{-2})$). The ratio ${\cal R}$ at $Q_T=0$ falls off approximately as $Q^{-0.85}$. The magnitude and the $Q$ dependence are in agreement with the newer results in \cite{Echevarria:2015uaa}, which includes higher order resummations and quark contributions (Fig.\ \ref{RQTatHiggs} (right)). Depending on the choice of $S_{NP}$ that is considered there, the fall-off at $Q_T=0$ varies from $Q^{-0.84}$ to $Q^{-1.1}$. 

In \cite{Boer:2014tka} not only the tail expressions (\ref{tailh1perp}) are considered, but also a model for TMDs that are approximately Gaussian at small $k_T$ and have the correct power-law tail behavior at large $k_T$. The difference with the "tail only" expressions using (\ref{tailh1perp}) is effectively a modification of $S_{NP}$. In \cite{Boer:2014tka} also the  dependence on the treatment of the very small $b$ region ($b<1/Q$, see next section) is studied. In Fig.\ \ref{RQTatHiggs} the range of predictions for ${\cal R}(Q_T)$ at $Q=m_H$ of \cite{Boer:2014tka} is shown and compared to that of \cite{Echevarria:2015uaa} where different $S_{NP}$ are considered and also the renormalization scale is varied. The conclusion is that the effect of linear gluon polarization in Higgs production is of order 2-5\%. The Higgs mass scale is thus sufficiently large to allow a reasonably precise prediction of the effect of linearly polarized gluons, even though the nonperturbative contributions are essentially unknown. 

Whether a few percent effect is measurable at LHC is another matter. It is smaller than the 10-20\% uncertainty of the perturbative NNLO+NNLL result of \cite{Neill:2015roa}. Besides that, the current $Q_T$ resolution of the Higgs is too low at low $Q_T$ (see e.g.\ ATLAS-CONF-2013-072), which ultimately will be somewhere around 5 GeV (private communication with Paolo Gunnellini, Hannes Jung and Pierre Van Mechelen). Moreover, depending on the Higgs decay channel, there may be different sources of dilution, such as due to the energy scale resolution (e.g.\ $\Delta Q \approx 0.5$ GeV in the $\gamma\gamma$ decay channel), which leads to smearing, and the background processes to deal with in numerator and denominator of ${\cal R}(Q_T)$. E.g.\ linearly polarized gluons enter in the process $gg \to \gamma\gamma$ without Higgs \cite{Nadolsky:2007ba,Qiu:2011ai}, although this is only a small contribution (a sub-percent level ${\cal R}(Q_T)$ at RHIC energy $\sqrt{s}=500$ GeV at low $Q$ ($4 < Q^2 < 30\ {\rm GeV}^2$) and low $Q_T$ ($0\leq Q_T \leq 1$ GeV) \cite{Qiu:2011ai}). Finally, at small $k_T$ $gg \to \gamma\gamma g$ needs to be included and could even dominate \cite{Szczurek:2014mwa}. As a further complication that channel is sensitive to different initial and final state interactions (ISI/FSI), cf.\ the study of linearly polarized gluons in Higgs plus jet production \cite{Boer:2014lka}. In conclusion, measuring the effect of linear gluon polarization in Higgs production at LHC may turn out to be too challenging.  

\section{Quarkonium production}
\begin{figure}[htb]
\centerline{
\includegraphics[width=5 cm]{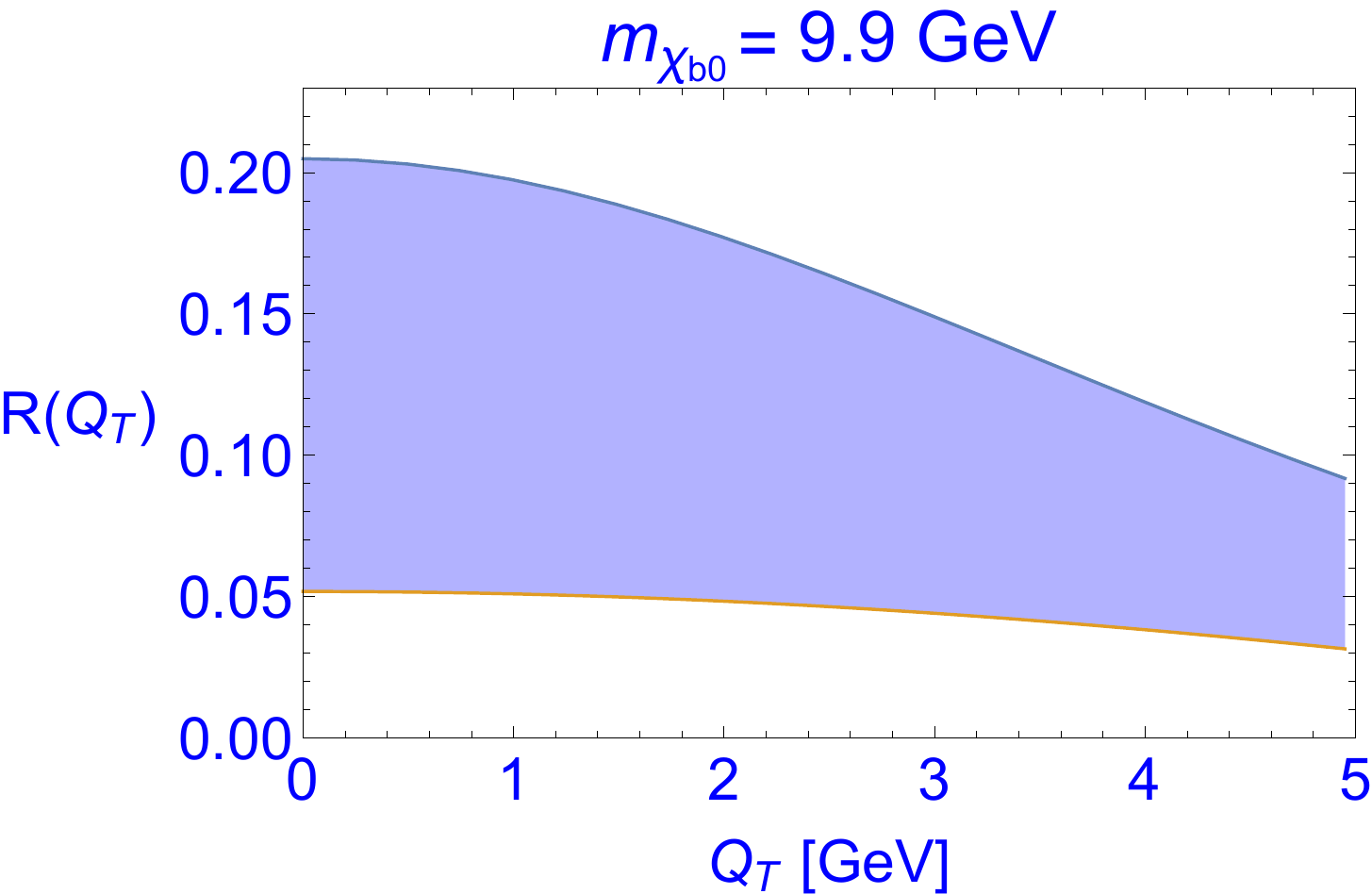}\hspace{1 mm}
\includegraphics[width=5 cm]{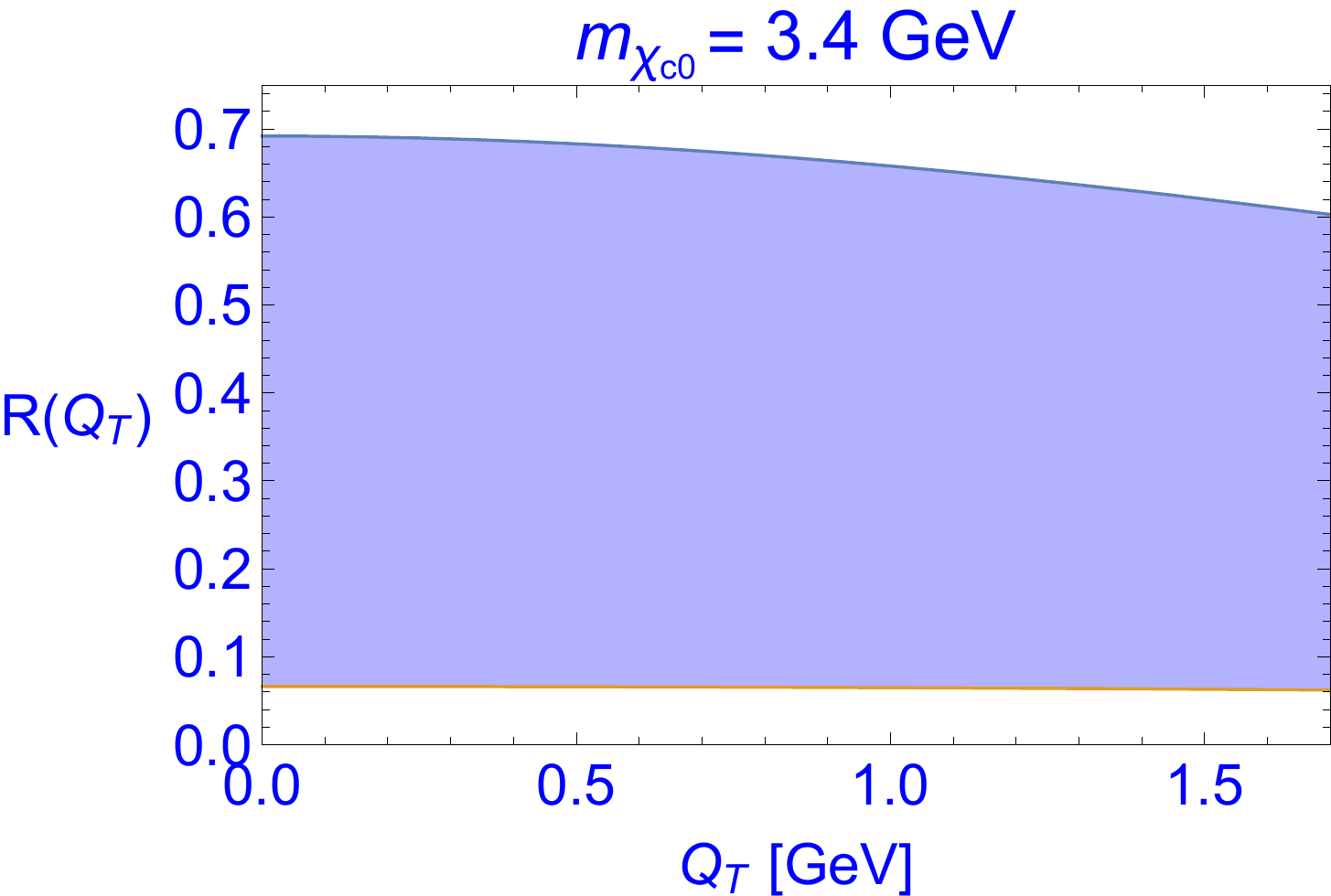}\hspace{1 mm}
\includegraphics[width=5 cm]{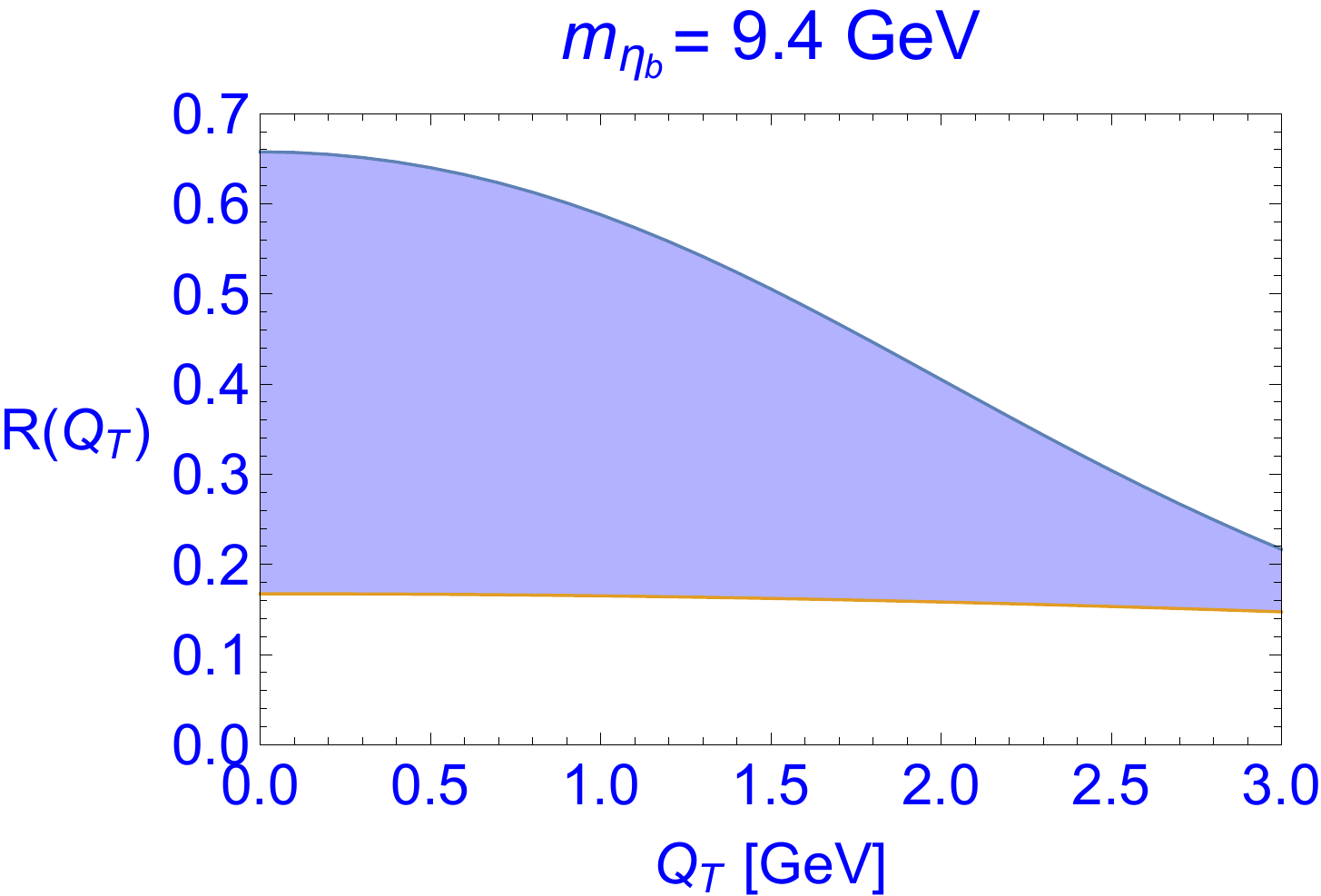}
}
\caption{\label{RQTquarkonia}Range of predictions of ${\cal R}(Q_T)$ for $\chi_{b0}$ (left) and $\chi_{c0}$ (middle) production from \cite{Boer:2014tka} and for $\eta_b$ production (right) from \cite{Echevarria:2015uaa}.} 
\end{figure}

Measuring the effect of linear gluon polarization in heavy (pseudo-)scalar $C=+$ quarkonium production seems more promising\footnote{In $J\neq 0$ quarkonium production, including $J/\psi$ and $\Upsilon$ production, the effects are either absent or suppressed.}, but here the theoretical uncertainties are significantly larger, preventing accurate predictions of the size of the effects. The ranges of predictions for ${\cal R}(Q_T)$ of \cite{Boer:2014tka} and \cite{Echevarria:2015uaa} shown in Fig.\ \ref{RQTquarkonia} correspond to the same variations as for Fig.\ \ref{RQTatHiggs}. Clearly large effects are possible in quarkonium production, but there are very large uncertainties. The increasing uncertainties with decreasing $Q$ as seen in the comparison of $Q=m_{\chi_{c0}}=3.4$ GeV and $Q=m_{\chi_{b0}}=9.9$ GeV, arises mainly from the treatment of the very small $b$ region, as will be discussed next.   

\subsection{Very small b region}
At low $Q$ there is quite some uncertainty from the very small $b$ region ($b <1/Q$) as displayed in Fig.\ \ref{RQTquarkonia}. 
In this section we will discuss this in more detail. First we consider the integral over all transverse momenta which formally corresponds
to $b=0$, i.e.\ if one integrates $W(q_T) \equiv {\rm F.T.}[\widetilde W(b)] $, which is the first term in Eq.\ (\ref{diffxs}), over all $\bm{q}_T$ one obtains $\widetilde W(0)$.
The perturbative expression for $S_A$ has the property that $e^{-S_A(0,Q)} = 0$, which forces the whole expression for $\widetilde W(0)$ to be zero 
and which will lead to a negative $W(q_T)$ for large $q_T$, outside its range of applicability, where the $Y$ term needs to be taken into account. This well-known (see e.g.\ \cite{Boglione:2014oea}) behavior of $W$ can be avoided by introducing the following regularization in the perturbative Sudakov factor $S_A$: $Q^2/\mu_b^2= b^2Q^2/b_0^2 \to Q^2/\mu_b^{\prime \, 2} \equiv b^2Q^2/b_0^2+1$ \cite{ParisiPetronzio}, which below is referred to as the `Parisi-Petronzio (PP) method'. 
It avoids the appearance of an infinitely large scale in $S_A$ as $b\to 0$. 
The precise form of the PP regularization usually is not very relevant since matching to the $Y$-term at large $q_T$ is needed anyway \cite{Bozzi:2005wk}. The scalar particle production case is special because the sensitivity to the very small $b$ region ($b < 1/Q$) already arises at $q_T=0$! 
But the region $b<1/Q$ at low $Q_T$ region ought to be power suppressed, 
that is of order $Q_T^2/Q^2$.
As it turns out (see below) the problem resides in the denominator of ${\cal R}$ that receives an increasing contribution from the $b<1/Q$ region as $Q$ decreases. The origin of the problem is however not with $S_A$ at $b= 0$, such that the PP method does not solve the problem, but rather it is with the unpolarized collinear gluon distribution at large scales and small $x$ values\footnote{I thank Markus Diehl for pointing this out.}. As $b$ decreases and $\mu_b$ increases, the latter grows without bound: $f_{g/P}(x; \mu_b)  \stackrel{b\to 0}{\longrightarrow} \infty$, while $\alpha_s(\mu_b) \stackrel{b\to 0}{\longrightarrow} 0$. Despite the fact that the integration over the small $b$ region remains finite, even if no form of regularization is introduced (indicated by `No regulator' below), the rapid growth of the unpolarized collinear gluon distribution especially at small $x$ causes the large dependence on the treatment of the very small $b$ region at smaller $Q$, as displayed in Figs.\ \ref{RQTatHiggs} (left) and \ref{RQTquarkonia} (left \& middle) for which $x=Q/\sqrt{s}$ with $\sqrt{s}=8$ TeV. 
\begin{figure}[htb]
\centerline{
\includegraphics[width=7.5 cm]{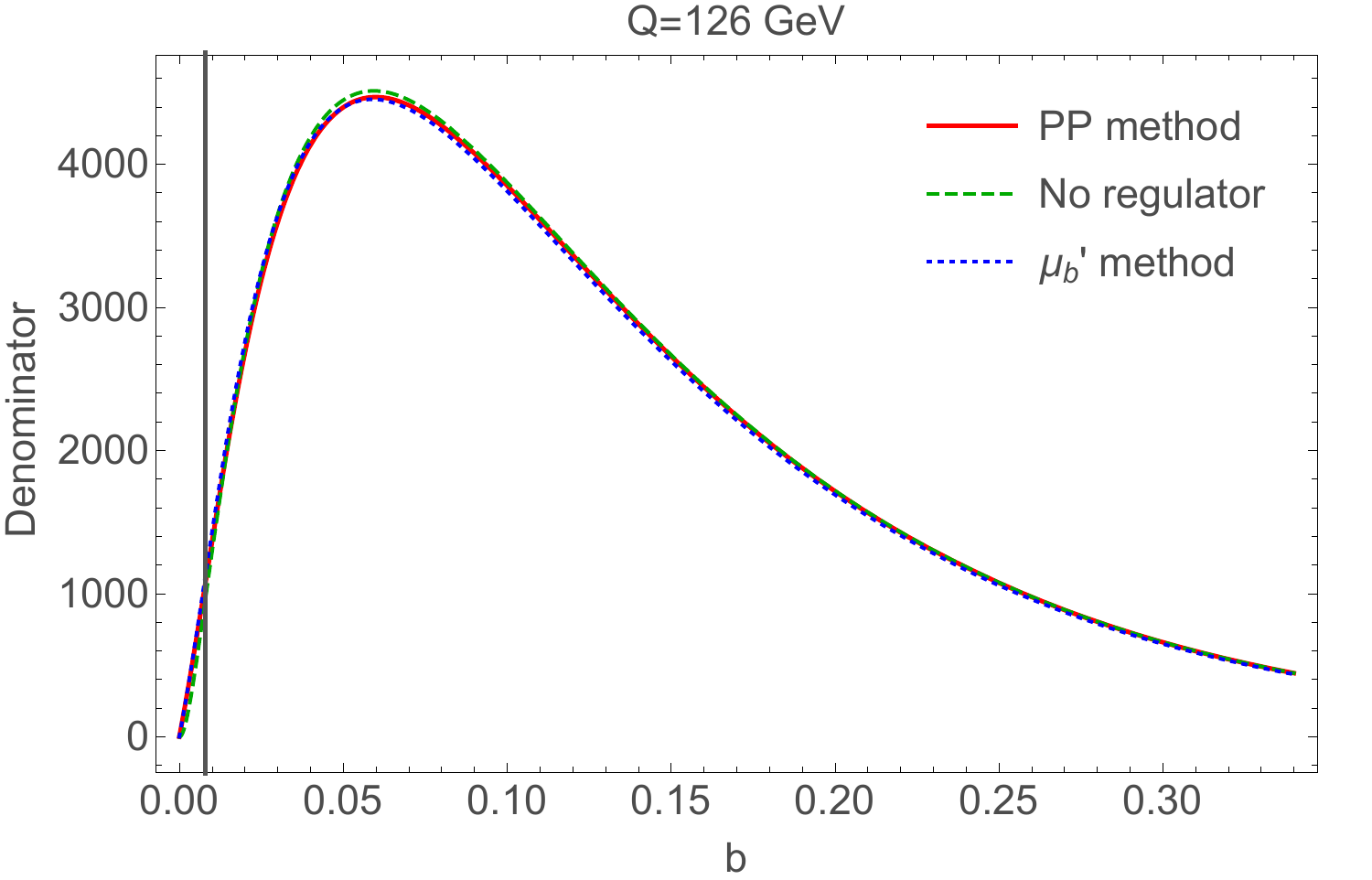}\hspace{1 mm}
\includegraphics[width=7.3 cm]{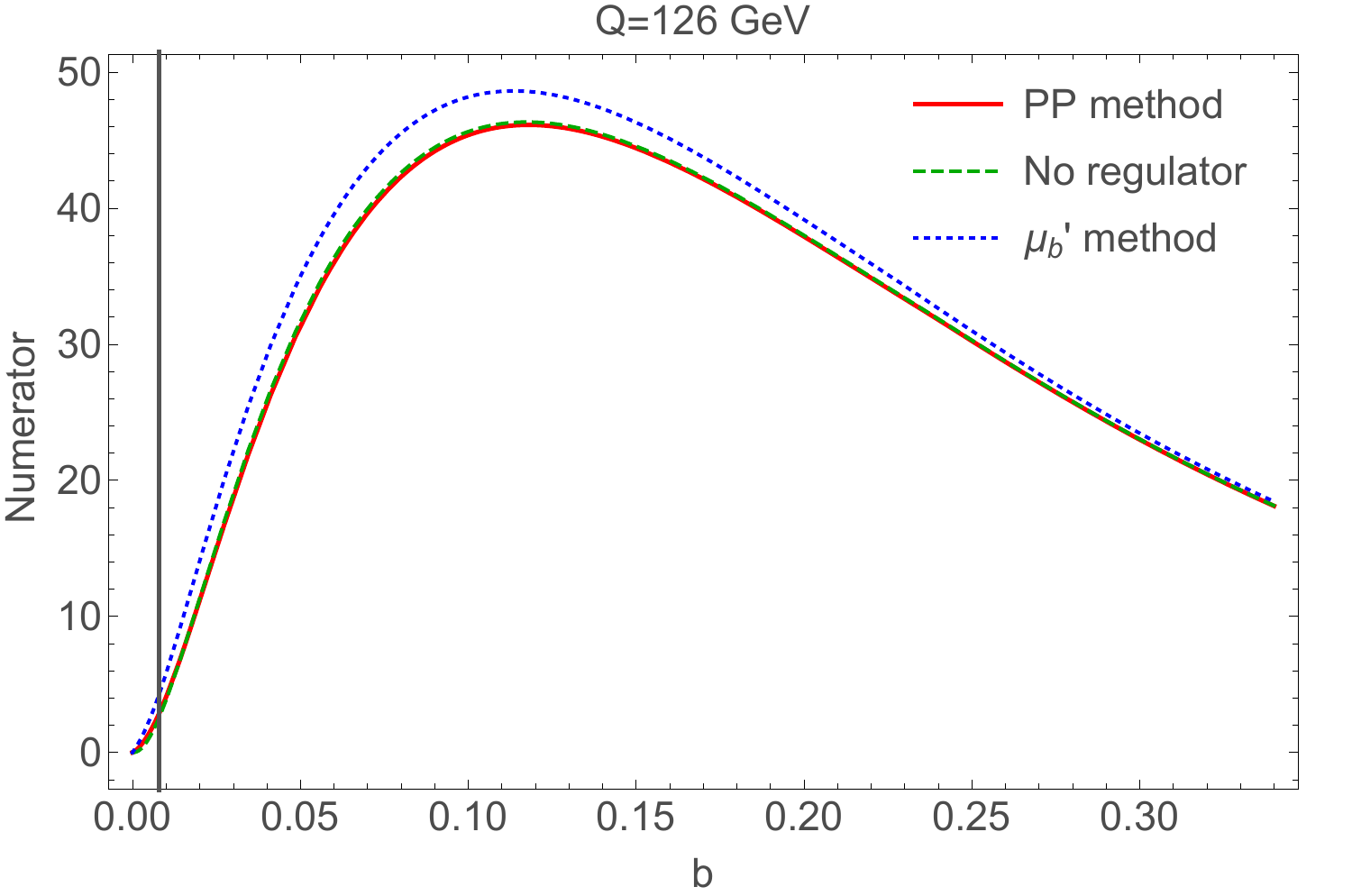}
}
\caption{\label{Ints126}Integrands $b \, \widetilde W(b)$ for the denominator and numerator of ${\cal R}$ at $Q=126$ GeV. The vertical line indicates where $b=1/Q$.} 
\end{figure}
\begin{figure}[htb]
\centerline{
\includegraphics[width=7.5 cm]{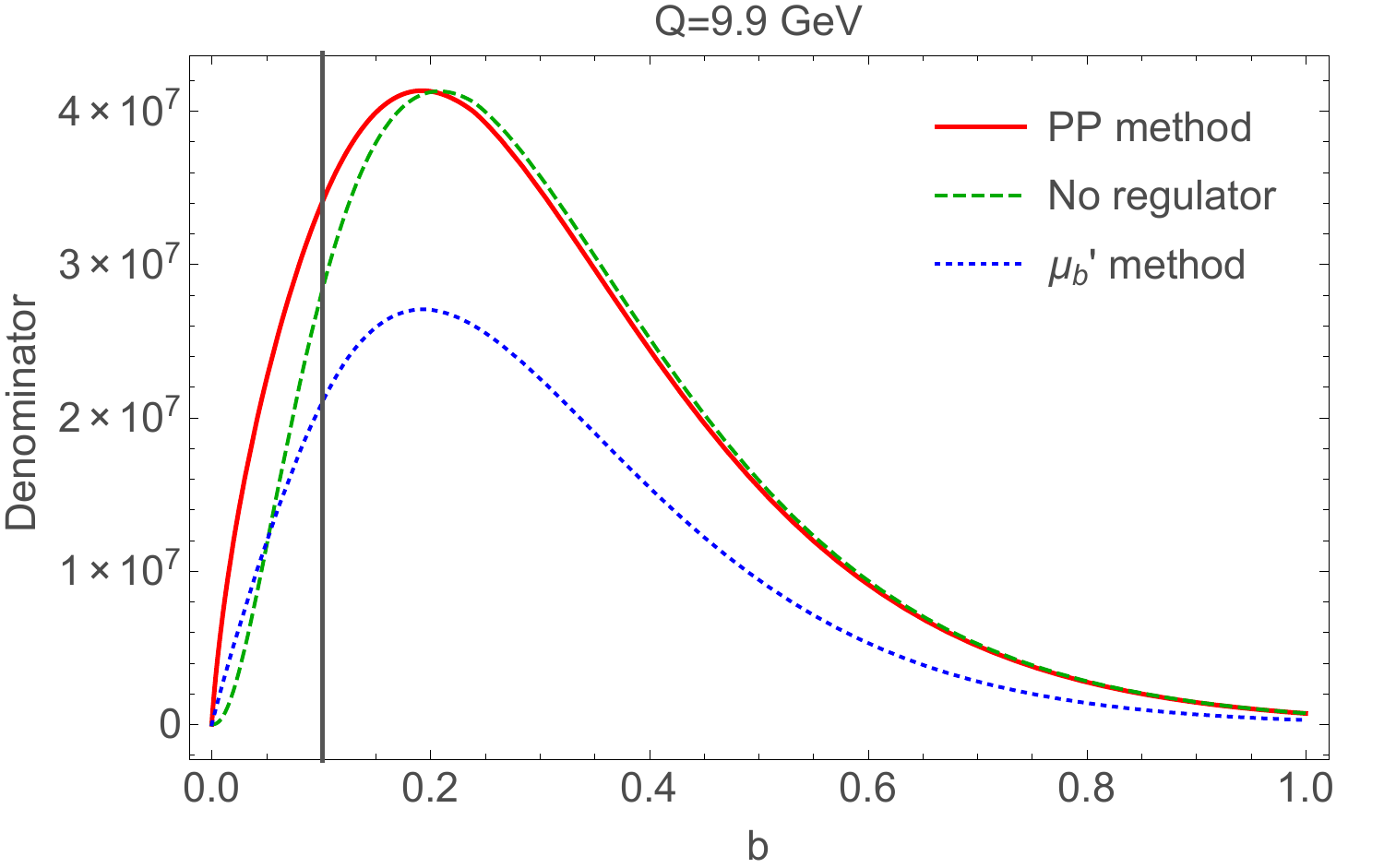}\hspace{1 mm}
\includegraphics[width=7.5 cm]{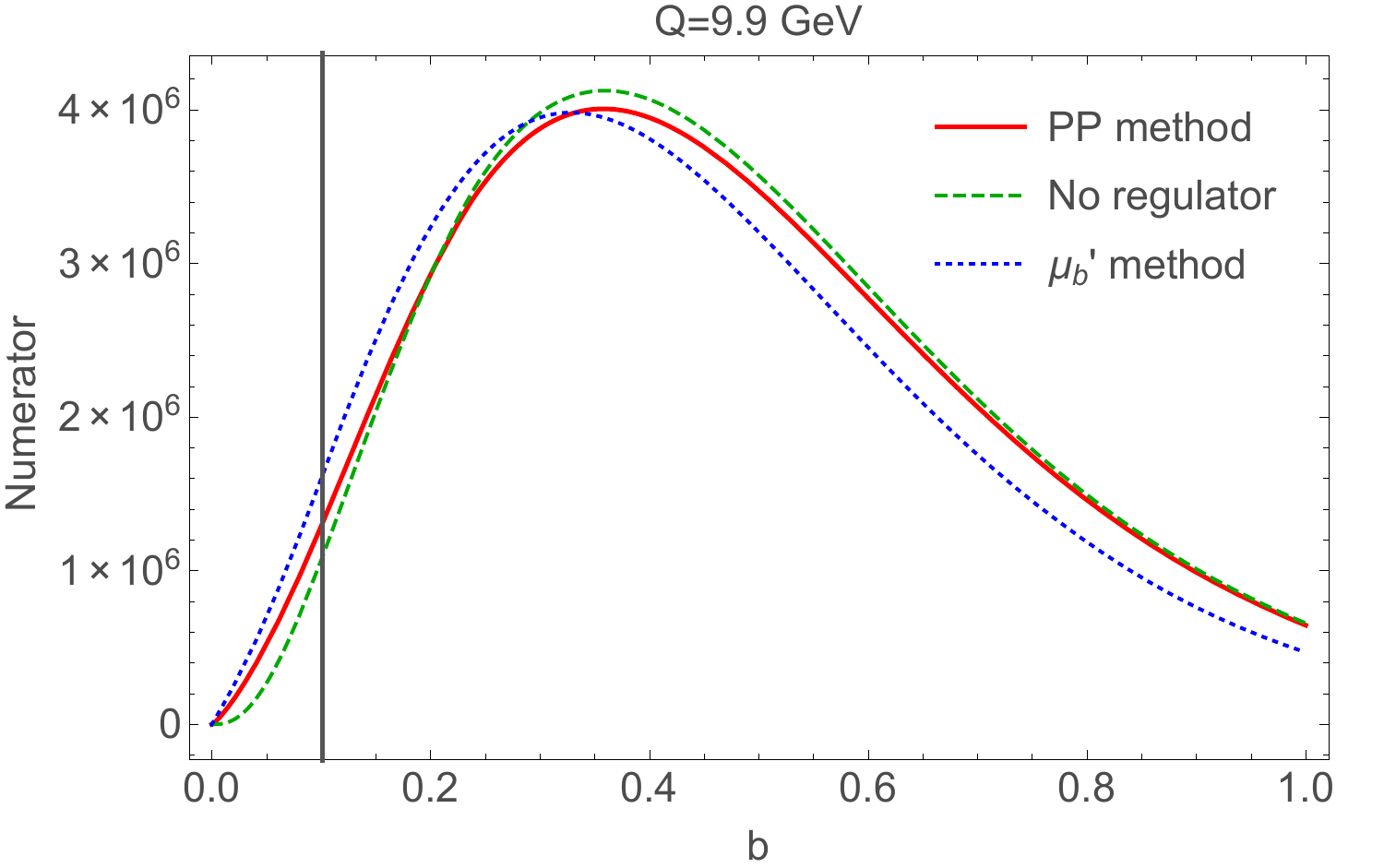}\hspace{1 mm}
}
\caption{\label{Ints9p9}Same as Fig.\ 3 but for $Q=9.9$ GeV.} 
\end{figure}
\begin{figure}[htb]
\centerline{
\hspace{1 mm}
\includegraphics[width=7.2 cm]{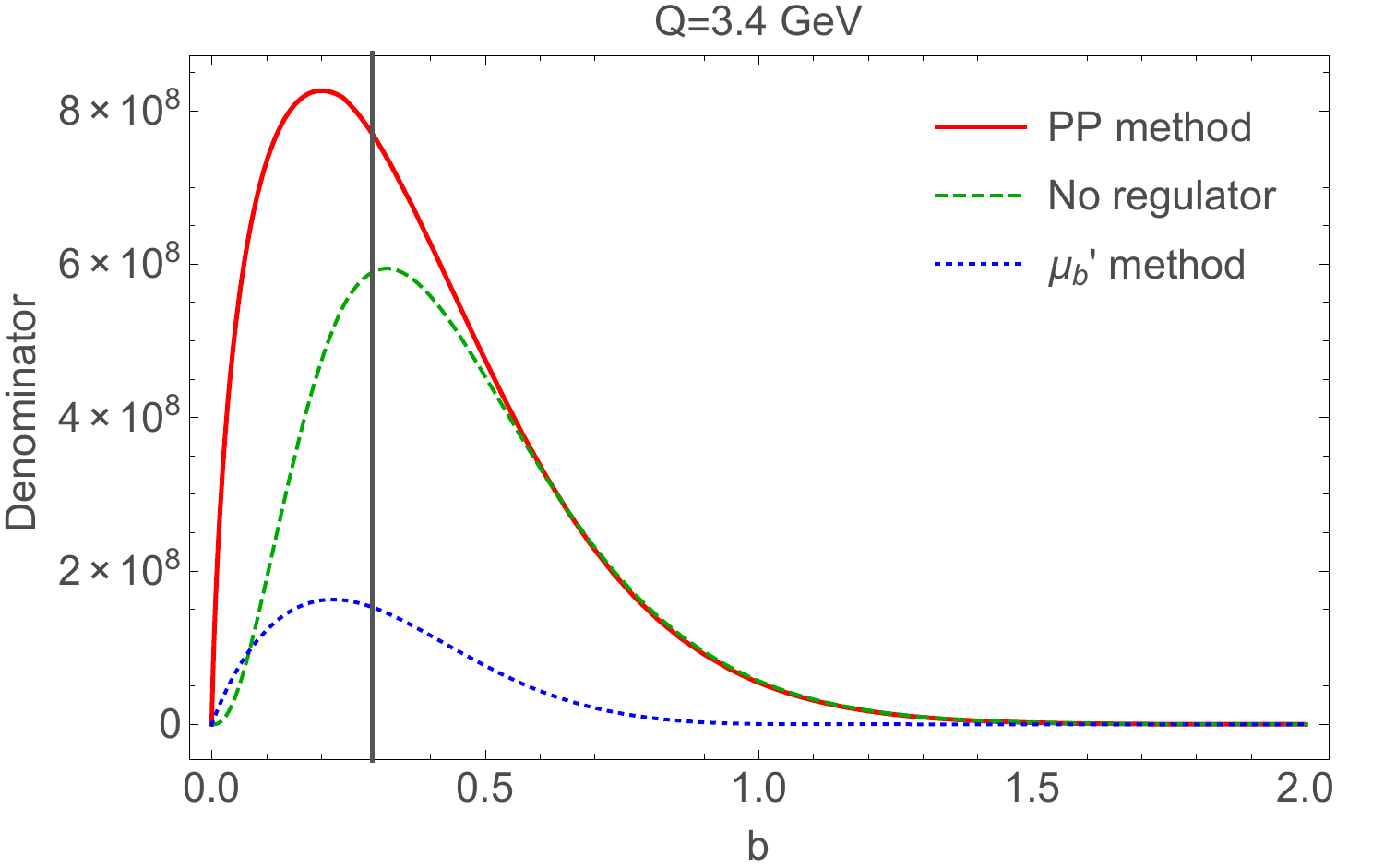}
\includegraphics[width=7.5 cm]{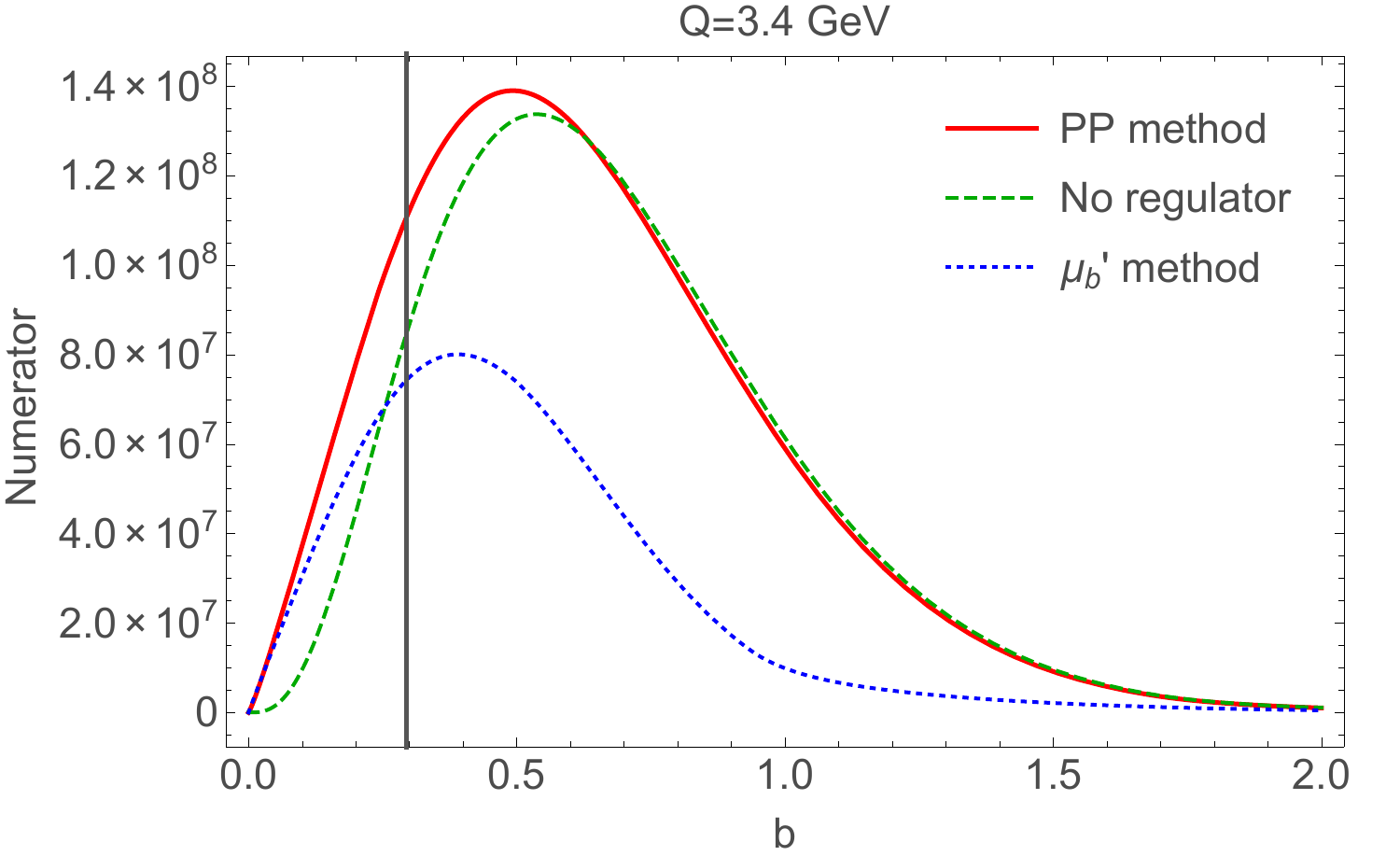}
}
\caption{\label{Ints3p4}Same as Fig.\ 3 but for $Q=3.4$ GeV.} 
\end{figure}
Figs.\ \ref{Ints126}-\ref{Ints3p4} show $b$ times $\widetilde W(b)$ for the denominator and numerator of ${\cal R}$ for three energies ($Q=126, 9.9, 3.4$ GeV) for three different treatments of the $b<1/Q$ region (again at $x=Q/\sqrt{s}$ with $\sqrt{s}=8$ TeV). Besides the unregulated `No regulator' expressions and the one with $S_A$ regulated with the PP method, we employ another way of regulating, referred to as the `$\mu_b^\prime$ method'. In this case the TMDs are evolved to the bounded scale $\mu_b^\prime \equiv [b^2/b_0^2+1/Q^2]^{-1/2}$  \cite{Boer:2014tka}\footnote{In \cite{Boer:2014tka} the scale $\mu_b^\prime$ was actually taken to be $[b/b_0+1/Q]^{-1}$, which for the present purpose would be equally fine. For the curves in that case, see the slides of the actual presentation at QCD evolution 2015.}, that is approximately $\mu_b$ for $b \gg 1/Q$ and $Q$ for $b \ll 1/Q$,
rather than to the unbounded $\mu_b$. This `$\mu_b^\prime$ method' tames the large rise of the unpolarized collinear gluon distribution, while at the same time regulates the behavior of $S_A$ as $b \to 0$. As can be seen the largest effect of the $\mu_b^\prime$ regularization arises in $\widetilde W(b)$ of the denominator at the lowest of the three energies, $Q=3.4$ GeV (Fig.\ \ref{Ints3p4} (left)), which corresponds to the smallest $x$ considered here ($x \approx 4 \cdot 10^{-4}$). The `$\mu_b^\prime$ method' seems a more appropriate way of regulating and leads to the largest ${\cal R}(Q_T)$ in Fig.\ \ref{RQTquarkonia} (left \& middle), which is very promising for measurements in quarkonium production. 
  
\subsection{Bottomonia}
Bottomonia have the advantage that they are less sensitive to the very small $b$ region than charmonia and the effect of linearly polarized gluons is likely not as small as in Higgs production. 
Using the color singlet model and NRQCD results, in \cite{Boer:2012bt} the following expressions are obtained for the differential cross sections of $C=+$ bottomonium production:
\begin{eqnarray}
\frac{d\sigma(\eta_b)}{d y\,d^{2}\mathbf{q}_T} & = & \frac{2}{9} \frac{\pi^3\alpha_s^2}{M^3\,s}\, \langle 0\vert {\cal{O}}^{\eta_b}_1 (^1 S_0)\vert 0\rangle\,  \mathcal{C}\left[f_{1}^{g}\, f_{1}^{g}\right]\,\left [1 - {\cal R}(\mathbf q_T^2) \right], \\[2 mm] 
\frac{d\sigma (\chi_{b 0})}{d y\,d^{2}\mathbf{q}_T} & = & \frac{8}{3} \frac{\pi^3\alpha_s^2}{ M^5\,s}\, \langle 0\vert {\cal{O}}^{\chi_{b0}}_1 (^3 P_0)\vert 0\rangle\,\mathcal{C}\left[f_{1}^{g}\, f_{1}^{g}\right]\, \left [ 1 + {\cal R}(\mathbf q_T^2)\right],\\[2 mm] 
\frac{d\sigma (\chi_{b 2})}{d y\,d^{2}\mathbf{q}_T} & =  & \frac{32}{9} \frac{\pi^3\alpha_s^2}{ M^5\,s}\, \langle 0\vert {\cal{O}}^{\chi_{b2}}_1 (^3 P_2)\vert 0\rangle \,\mathcal{C}\left[f_{1}^{g}\, f_{1}^{g}\right] ,
\end{eqnarray}
with the same ${\cal R}(\mathbf q_T^2)$ as in Eq.\ (\ref{calRQT}). The hadronic matrix elements $\langle 0\vert {\cal{O}}^{[\overline Q Q]}_1 \vert 0\rangle$, and hence the uncertainties about the hadronic wave functions, drop out when normalizing to the integrated cross section: 
\begin{eqnarray}
\frac{1}{\sigma(\eta_b)} \frac{d\sigma(\eta_b)}{d y\,d^{2}\mathbf{q}_T} & = & \frac{\mathcal{C}\left[f_{1}^{g}\, f_{1}^{g}\right]}{f_{1}^{g}\, f_{1}^{g}}\,\left [1 - {\cal R}(\mathbf q_T^2) \right],\\[2 mm] 
\frac{1}{\sigma (\chi_{b 0})} \frac{d\sigma (\chi_{b 0})}{d y\,d^{2}\mathbf{q}_T} & = & \frac{\mathcal{C}\left[f_{1}^{g}\, f_{1}^{g}\right]}{f_{1}^{g}\, f_{1}^{g}}\, \left [ 1 + {\cal R}(\mathbf q_T^2)\right],\\[2 mm] 
\frac{1}{\sigma (\chi_{b 2})} \frac{d\sigma (\chi_{b 2})}{d y\,d^{2}\mathbf{q}_T} & =  & \frac{\mathcal{C}\left[f_{1}^{g}\, f_{1}^{g}\right]}{f_{1}^{g}\, f_{1}^{g}}.
\end{eqnarray}
Extracting ${\cal R}(Q_T)$ for a single state is difficult, as it is a modification of a cross section contribution that cannot be isolated separately. However, by measuring the cross section for two or more bottomonia it may become feasible to probe ${\cal R}(Q_T)$ directly, by considering ratios of ratios: 
\begin{eqnarray}
\frac{\sigma (\chi_{b 2})}{\sigma(\chi_{b 0})} \frac{d\sigma(\chi_{b 0})/d^2 \mathbf{q}_T}{d\sigma (\chi_{b 2})/d^2 \mathbf{q}_T} & \approx & 1 + {\cal R}(\mathbf q_T^2), \label{ratioofratio1}\\
\frac{\sigma (\chi_{b 0})}{\sigma(\eta_b)} \frac{d\sigma(\eta_b)/d^2 \mathbf{q}_T}{d\sigma (\chi_{b 0})/d^2 \mathbf{q}_T} & \approx & \frac{1 - {\cal R}(\mathbf q_T^2)}{1 + {\cal R}(\mathbf q_T^2)}.\label{ratioofratio2}
\end{eqnarray}
Because of the very small scale differences, $m_{\eta_b} \approx m_{\chi_{b0}} \approx m_{\chi_{b2}} $, evolution will not play a significant role in the comparison. Consistency between the results for (\ref{ratioofratio1}) and (\ref{ratioofratio2}) serves as a cross-check. 

The above expressions are all color singlet model expressions \cite{Baier:1983va}, which may be justified
for $C=+$ bottomonium states from NRQCD considerations and by several numerical studies of color octet contributions \cite{Bodwin:1994jh,Hagler:2000dd,Maltoni:2004hv,Bodwin:2005hm,Lansberg:2012kf}.
The measurement of the transverse momentum distribution of such bottomonium states will not be easy, but is in principle possible at LHC. 

It should be mentioned that there are other processes involving heavy quarks, that could probe the linear gluon polarization directly. These exploit angular modulations of the cross section. The best opportunity may be offered by open charm or bottom production in electron-proton collisions: $e\, p \to Q\, \overline{Q}\, X$. Here $h_1^{\perp \, g} $ appears by itself, rather than in a product of two, so the effect should be less suppressed and moreover, the sign of the function can be determined. Nonzero $h_1^{\perp \, g} $ leads to $\cos 2\phi$ asymmetries in heavy quark pair electro-production \cite{Boer:2010zf}, which are maximally 15\% asymmetries \cite{Pisano:2013cya} and are best measured at an Electron-Ion Collider (EIC). In contrast, the analogous process $p\, p \to Q\, \overline{Q}\, X$ at RHIC or LHC runs into the problem of TMD factorization breaking \cite{Rogers:2010dm}.

\section{Small-x aspects}
A natural question is whether the linear gluon polarization plays a role at small $x$. It is known that the distribution of circularly polarized gluons $\Delta g(x)$ is suppressed w.r.t.\ $g(x)=f_{g/P}(x)$ at small $x$, because its evolution kernel does not have a $1/x$ behavior. In contrast, the perturbative tail of the linearly polarized gluon distribution inside unpolarized protons does grow with $1/x$ and is driven by the unpolarized gluon distribution, cf.\ Eq.\ (\ref{tailh1perp}). There is no theoretical reason why the amount of linear polarization should be small at small $x$. It turns out the linear polarization can even become maximal at small $x$. E.g.\ the small-$x$ ``$k_T$-factorization'' approach involves maximum polarization \cite{Catani:1990eg}, i.e.\
\[
 \Gamma_g^{\mu\nu}(x,\bm{k}_T )_{\rm max\ pol}= 
\frac{2}{x}\,\frac{k_T^\mu k_T^\nu}{\bm{k}_T^2}\,f_1^g(x,\bm{k}_T^2).
\]
This expression has been used in the study of Higgs production in \cite{Lipatov:2014mja}.

It is expected that at very small $x$ gluon saturation will take place. The effect of saturation has been studied in the CGC framework of the McLerran-Venugopalan (MV) model. 
It was recognized in \cite{Dominguez:2011wm} that at small-$x$ (and large $N_c$) there are two distinct unpolarized gluon distributions to consider: 
the Weizs\"acker-Williams (WW) and the dipole (DP) distribution (elucidating ``A tale of two gluon distributions'' by Kharzeev, Kovchegov \& Tuchin \cite{Kharzeev:2003wz}). The WW and DP distributions of unpolarized and linearly polarized gluons have been studied within the MV model \cite{Metz:2011wb}, showing that the dipole $h_1^{\perp \, g}$ distribution is maximal for all transverse momenta, whereas the WW $h_1^{\perp \, g}$ is suppressed w.r.t.\ $f_1^g$ in the saturation region ($k_T \ll Q_s$) and maximal outside it ($k_T \gg Q_s$).  

Which distribution (WW or DP) is of relevance depends on the considered process. Ref.\ \cite{Dominguez:2011wm} lists processes that probe WW and/or DP distribution for unpolarized gluons. Not all of these allow a study of $h_1^{\perp \, g}$. It turns out that DIS, DY, SIDIS, hadron and $\gamma+{\rm jet}$ production in $pA$ collisions are in leading power not sensitive to $h_1^{\perp \, g}$ \cite{Boer:2009nc}. In dijet production in $ep$ and $eA$ collisions at small $x$ the WW distribution is probed and in dijet production in $pp$ and $pA$ collisions at small $x$ a combination of the WW and DP distributions are probed. The latter holds in the large $N_c$ limit, otherwise additional gluon distributions enter. Although in general dijet production in $pp$ and $pA$ collisions suffers from factorization breaking effects \cite{Rogers:2010dm}, in the small-$x$ limit factorization breaking contributions may become suppressed, effectively allowing TMD factorization \cite{Chirilli:2011km,Kotko:2015ura}.  
     
Dijet production in $ep$ and $eA$ collisions can be studied at an EIC. Since the expectations for the WW $h_1^{\perp \, g}$ inside and outside the saturation region differ a lot, it would thus be very interesting to study dijet DIS at a high-energy EIC as a function of dijet transverse momentum and $x$. Expressions involving the $h_1^{\perp \, g}$ for general $x$ can be found in \cite{Pisano:2013cya}, while small-$x$ expressions can be found in \cite{Metz:2011wb,Dumitru:2015gaa}. 
We note that for dijet DIS these expressions involve only the WW-type distributions without requiring large $N_c$. 
As a side remark, dijet DIS also happens to be the golden channel for the gluon Sivers effect at EIC. Just as the gluon Sivers function would be zero without ISI/FSI, the {\it difference} between WW and DP distributions would be absent without ISI/FSI. 
  
\section{Summary}
In this contribution the effects of linear gluon polarization have been studied. Using TMD evolution (to different levels of accuracy) 
the effect in Higgs production is estimated to result in small contributions (2-5\% level) from linearly 
  polarized gluons at the Higgs mass scale. This is smaller than the current theoretical uncertainty in the perturbative cross section description (NNLL+NNLO) and given the poor $Q_T$-resolution below 10 GeV, the feasibility of measuring the effects of linear gluon polarization in Higgs production at the LHC is doubtful.
Studies of $C=+$ quarkonium states may offer promising alternatives, but here the predictions are very sensitive to the
  treatment of the very small-$b$ region and to the large-$b$ region. It leads to much larger effects, but unfortunately with much larger theoretical uncertainties.  
Future data from LHC on the bottomonium states $\chi_{b0/2}$ and $\eta_b$ are most promising.  The bottomonium scale may be optimal from the perspective of reducing contributions from the very small-$b$ and large-$b$ regions while maintaining sufficiently large effects from linear gluon polarization. In addition, evolution will hardly play a role in the comparison of different bottomonium states. Heavy quark pair and dijet production in DIS at a high-energy EIC may exhibit   
  large $h_1^{\perp g}$ effects too, allowing ways to study its sign, its small-$x$ behavior, and, even saturation effects. 


\end{document}